\documentclass{emulateapj}
\usepackage{textcomp}
\usepackage{mathtext}
\usepackage{amssymb}
\usepackage{rotating}
\usepackage{amsmath}

\newcommand{\fermi}{\textit{Fermi}}
\newcommand{\gr}{$\gamma$-ray}
\newcommand{\uG}{\ \mu{\rm G}}

\newcommand{\gray}{$\gamma$-ray}
\newcommand{\grays}{$\gamma$-rays}

\begin{document}

\title{{\it Fermi} observation of the jets of the microquasar SS433}

\author{Yi Xing\altaffilmark{1}, Zhongxiang Wang\altaffilmark{1}, Xiao Zhang\altaffilmark{2}, Yang Chen\altaffilmark{2,3}, V. Jithesh\altaffilmark{4}}

\altaffiltext{1}{Key Laboratory for Research in Galaxies and Cosmology, Shanghai Astronomical Observatory, Chinese Academy of Sciences,
80 Nandan Road, Shanghai 200030, China}

\altaffiltext{2}{School of Astronomy \& Space Science, Nanjing University, 163 Xinlin Avenue, Nanjing~210023, China}

\altaffiltext{3}{Key Laboratory of Modern Astronomy and Astrophysics, Nanjing University, Ministry of Education, China}

\altaffiltext{4}{Inter-University Centre for Astronomy and Astrophysics, Pune 411 007, India}

\begin{abstract}
We report our analysis of the data obtained with the Large Area Telescope (LAT)
onboard the {\it Fermi Gamma-ray Space Telescope (Fermi)} for 
the SS 433/W50 region.  The total data length is ten years.
Different from the previous results reported for this region (for which
an old version of database was used), we show that
excess emission is detected around the w1 region in the western lobe 
of the jets from SS 433. The 
region is bright at X-rays due to the interaction between the jet and
the ambient medium. This detection also matches well the recent results of
the very-high-energy detection of SS 433/W50 with the High Altitude Water
Cherenkov (HAWC). However, the eastern regions that are slightly brighter
in HAWC's observation are not detected in the {\it Fermi} data. 
Constructing the broad-band spectral
energy distribution (SED) for the western region, we compare with the HAWC 
results
for the eastern regions and discuss the possible origin of the emission.
A leptonic scenario can provide a fit to the {\it Fermi} GeV spectrum
and HAWC TeV detection, where the former and latter are due to the 
synchrotron radiation and inverse-Compton process respectively. However,
the model can not explain the X-ray and radio emission from the region 
simultaneously,
which thus requires further observational and theoretical studies
of the region in order to clarify the reasons.

\end{abstract}


\section{Introduction}

SS 433 is probably one of the most well-known sources in the sky that has been
extensively studied from radio to very high 
energies \citep{fab04,bor+15,kar+17,magic18}.
This binary contains
a compact object and an A-type evolved star \citep{ghm02}, with an orbital
inclination angle of 78\arcdeg\ \citep{eik+01}. Two jets from the compact
object point to the east and west 
directions, interacting with interstellar gas contained in a size of
$2\arcdeg\times 1\arcdeg$ supernova remnant W50.
The distance to SS 433/W50 is 5.5$\pm$0.2 kpc \citep{bb04}, implying
the physical size of $180\ {\rm pc}\times 80\ {\rm pc}$ for W50 
(e.g., \citealt{gbb11}).
Within W50, different features have been found 
(for a review, see, e.g., \citealt{fab04}).  In particular
at X-rays, two lobes extending from SS~433 to eastern and western regions of 
$\sim$60\arcmin\ away are clearly visible \citep{was+83,bak96}. 
Five bright regions (e1--e3 in the eastern lobe and w1 and w2
in the western lobe) were selected for detailed studies, from which these
regions were considered as the results of the interaction between the jets and
the ambient medium \citep{so97,sp99,bri+07}.

Recently the very-high-energy (VHE) emission (at $\sim$20 TeV) from 
the eastern and western interaction regions has been detected by 
the High Altitude Water Cherenkov (HAWC) observatory with a significance 
of 5.4$\sigma$ \citep{abe+18}. Modeling of the broadband spectral energy 
distribution (SED) of the e1 region indicates a leptonic scenario for
the TeV \gr\ emission, in
which a single population of electrons with energies up to hundreds of TeVs 
radiates the synchrotron emission at radio to X-ray frequencies and
inverse-Compton (IC) scatter cosmic microwave background (CMB)
photons to the observed TeV
ones \citep{abe+18}. The HAWC results have established again, in addition
to blazars that are the major class of \gray\ sources in the sky due to 
their jets, jets associated with Galactic stellar sources are also
strong high-energy \gray\ emitters.

However at GeV energies, only excess emission possibly associated with
SS 433/W50 was reported from analysis of data obtained with 
the Large Area Telescope (LAT) onboard the 
{\it Fermi Gamma-ray Space Telescope (Fermi;} \citealt{bor+15}).
This excess \gray\ emission at GeV energies was not clearly associated 
with the eastern or 
western interaction regions, different from that indicated by the HAWC 
detection. Also the LAT data analyzed in \citet{bor+15} were obtained
during the first 5.5 years of \fermi\ and selected from the P7rev\_v15 
database. Given these reasons, we re-analyzed the LAT data that were 10 years
long and from the latest Pass 8 database. We found different results and
report them in this paper. Below we describe the \fermi\ data and
source model for analysis in Section 2,
and present our analysis and results in Section 3.
In Section 4, the results are discussed by considering different 
physical processes.
\begin{figure*}
\centering
\epsscale{1.0}
\plottwo{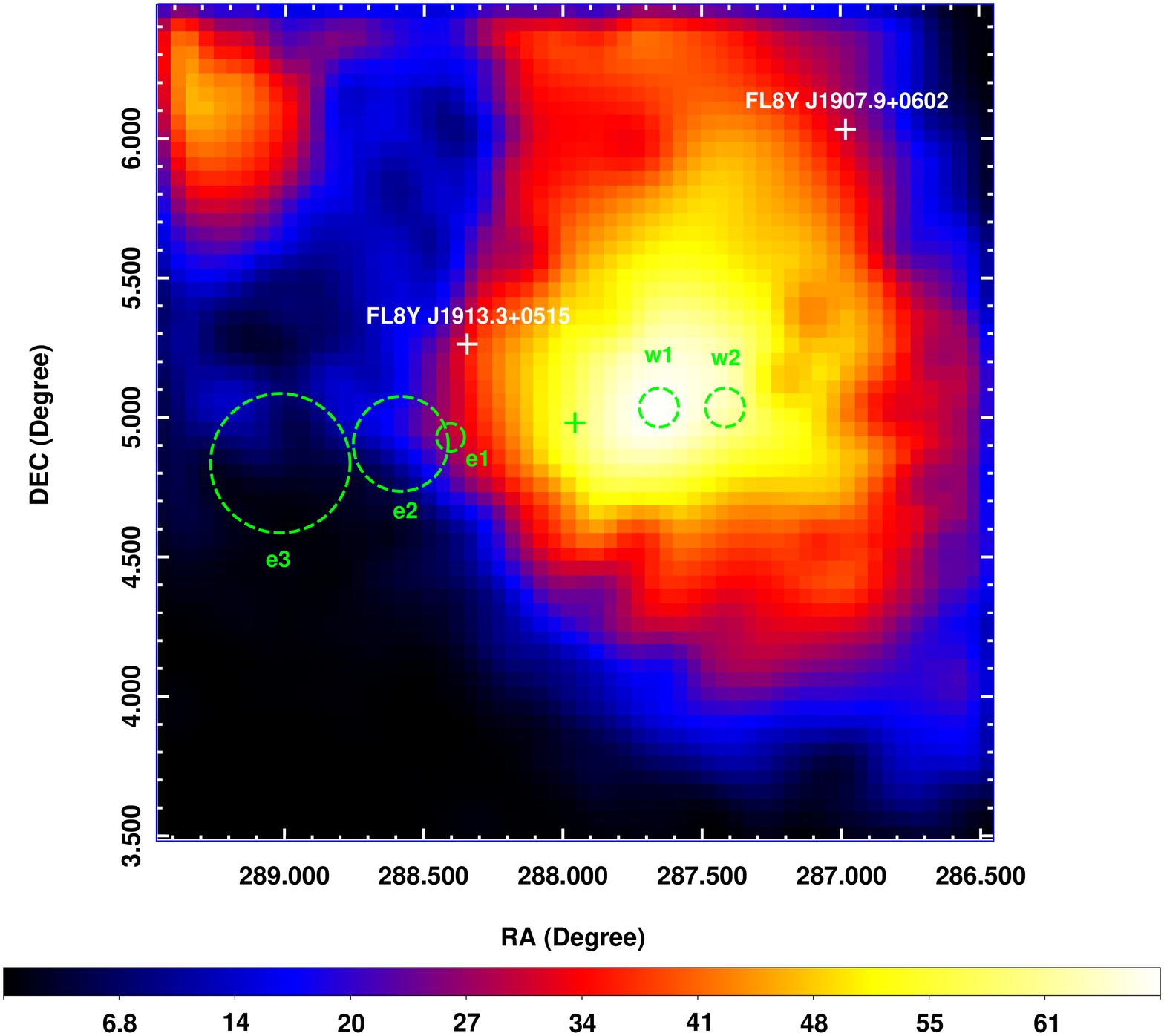}{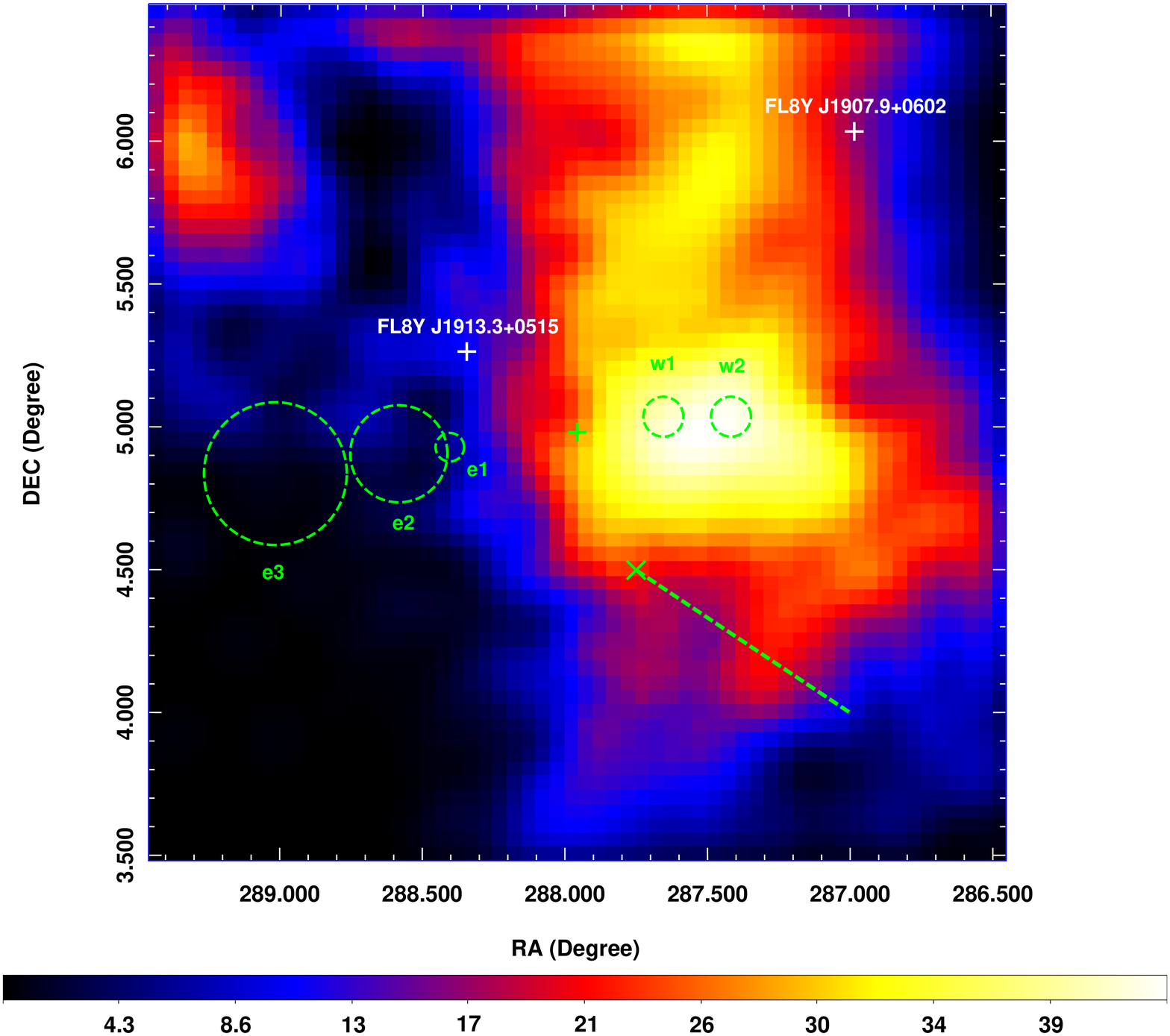}
\caption{$\mathrm{3^{o}\times3^{o}}$ TS maps in 0.3--100 GeV band, extracted 
using 10-year data ({\it left} panel) and the data of the same time period 
as that in \citet{bor+15} ({\it right} panel). The image scale of the maps
is 0\fdg05 pixel$^{-1}$. TS value ranges are indicated by the color bars.
The white plus signs mark the positions of the catalog sources within 
the region. 
The green plus signs mark the position of SS 433 and the green circles mark the 
bright X-ray emission regions of e1--e3, and w1 and w2 defined
from X-ray observations \citep{so97}. In the right panel for a 
comparison, the high-TS region shown in Figure 1 of \citet{bor+15} is indicated 
by the green cross sign and dashed line, which is far off the w1 and w2 
regions.}
\label{fig:tsmap}
\end{figure*}

\section{LAT Data and Source Model}
\label{sec:obs}

LAT is a $\gamma$-ray imaging instrument scanning the whole sky every 
three hours \citep{atw+09}. In the analysis, we selected 10 years 
(from 2008-08-04 15:43:36 UTC to 2018-08-05 00:00:00 UTC) LAT events 
in the 0.1--100 GeV range from the \textit{Fermi} Pass 8 database.
A $\mathrm{20^{o}\times20^{o}}$ region centered at the position of SS 433, 
which is R.A.=19$^{\rm h}$11$^{\rm m}$49$\fs$56 and
Decl.=$+$04$^{\circ}$58$'$57$\farcs$8, were considered.
We included the LAT events with zenith angles less than 90 degrees to prevent 
the Earth's limb contamination, and excluded the events with quality flags 
of `bad'. Both these selections are recommended by the LAT 
team\footnote{\footnotesize http://fermi.gsfc.nasa.gov/ssc/data/analysis/scitools/}.

We used the \textit{Fermi} LAT 4-year catalog (3FGL, \citealt{3fgl15}) and 
the preliminary LAT 8-year point source list\footnote{\footnotesize https://fermi.gsfc.nasa.gov/ssc/data/access/lat/fl8y/} 
(FL8Y) to make the source model. 
Sources within 5 degrees from SS 433 in the FL8Y were included in 
the source model and their spectral parameters were set free; 
sources 5--20 degrees away from SS 433 in 3FGL 
were included with their spectral parameters fixed to the values given 
in the catalog.  The spectral forms of these sources are provided 
in the two catalogs. We note that FL8Y, released in early 2018, is not 
encouraged to be used directly and the extended source templates and 
the Galactic/extragalactic diffuse emission models have not been updated 
accordingly.
Therefore we only used the sources within 5 degrees from SS 433 in FL8Y.
In addition, the background Galactic and extragalactic diffuse 
emission were included by adding the spectral model 
gll\_iem\_v06.fits and the file iso\_P8R2\_SOURCE\_V6\_v06.txt, respectively,
to the source model.
The normalizations of the diffuse components were set as
free parameters.

\section{Data Analysis and Results} 
\label{sec:ana}

\subsection{Likelihood Analysis}
\label{subsec:la}

We performed standard binned likelihood analysis to the LAT data in 
the 0.3--100 GeV band. The version of the LAT science tools software package 
used was {\tt v11r5p3}.
LAT events below 300 MeV were not included because of the relatively 
large uncertainties of the instrument response function of the LAT and 
strong background emission from the Galactic plane in the low energy range. 
A $\mathrm{3^{o}\times3^{o}}$ residual Test Statistic (TS) map 
in 0.3--100 GeV band centered at SS 433 was constructed (left panel 
of Figure~\ref{fig:tsmap}), in which all the known sources given
in the LAT catalogs were removed.
The TS value is a measurement of the fit improvement for including 
a source at a position in the source model, and is approximately 
the square of the detection significance of the source \citep{1fgl}. 
It can be seen from the TS map that significant excess emission is 
detected around the western region w1,
as the maximum TS value is $\sim$65 around w1 
(corresponding to $>$7$\sigma$ detection significance).
In the eastern region, TS$\sim$30 around e1 but the non-negligible value
is likely due to emission in the western region; note that the 68\% containment
angle of LAT is $\sim$2 degrees at 300 MeV.
In this analysis, we found photon index $\Gamma= 6.0\pm$1.3 
and 0.3--100 GeV flux
$F_{0.3-100}= 8.4\pm1.3\times 10^{-9}$ photons~s$^{-1}$\,cm$^{-2}$ for w1, 
indicating a very steep source spectrum.

\citet{bor+15} has reported the detection of \gray\ emission toward 
the SS 433/W50 region. However in their TS map, the excess emission was in
the south west of SS433, far off the w1 and w2 X-ray lobe,
and moreover it does not exist in our TS map (see Figure~\ref{fig:tsmap}). 
The differences in data analysis between theirs and this work are that
they used the P7rev\_v15 database and the time period of the data was 
the first 5.5 years (MJD 54682.6--56719.4) of the \fermi\ observation. 
In order to check the reasons
for the differences, we repeated their analysis but used the Pass 8 data.
The obtained TS map is shown in the right panel of Figure~\ref{fig:tsmap}. 
As can be seen, the TS map is nearly the same as that from the 10 years data, 
while with a maximum value of $\sim$40 in the w1 and w2 regions.
We therefore conclude that the large differences are likely due to 
our use of
the updated version of the LAT data and instrument response functions, and
our results are more consistent with the general features of the region
obtained at X-ray and TeV energies. Because of these, we do not consider
the results in \citet{bor+15} in the following analysis and discussion.

In addition, for the purpose of also confirming that there is no detected 
emission in the
eastern region, we performed likelihood analysis to the LAT data of SS 433/W50
by including two point sources at the central positions of e1 and w1. 
A power law emission was assumed for them.
We found that significant \gr\ emission can only be detected at w1 
with a TS value of 64,
while no significant emission can be detected at e1,
as the TS value is $\sim$0. Nearly the same values of photon index and
flux in 0.3--100 GeV as the above treating w1 as a single source were obtained.

\subsection{Spectral Analysis}
\label{subsec:sa}

We extracted the \gr\ spectrum of the western region by performing maximum 
likelihood analysis of the LAT data in 12 evenly divided energy bands 
in logarithm from 0.1 GeV to 100 GeV. 
The spectral normalizations of the sources within 5 degrees from SS 433 were
set as free parameters, and all the other parameters of the sources in 
the source model were fixed at the values obtained from the above maximum 
likelihood analysis.
A power law model was assumed for the excess emission, with the photon index 
fixed at 2.
The obtained spectrum data are given in Table~1, and
the spectrum is shown in Figure~\ref{fig:sed}. 
Only those spectral data points with their flux values 2 times greater 
than the flux uncertainties are kept. For other data points, their
95\% (2$\sigma$) flux upper limits were derived and given.
It can be seen that the \gr\ emission from the western region can only be 
detected in the energy ranges from $\sim$300 MeV to $1.8$ GeV and the flux
decreases steeply, consistent with the above result from
the maximum likelihood analysis.

For comparison, we also derived the 95\% flux upper limits for 
the e1 region and provide them in Table~1.
\begin{table}
\centering
\tabletypesize{\footnotesize}
\tablecolumns{10}
\tablewidth{240pt}
\setlength{\tabcolsep}{2pt}
\caption{\fermi\ LAT flux measurements and upper limits of the w1 and e1 regions}
\label{tab:spectra}
\begin{tabular}{lccccc}
\hline
$E$ & Band & $F/10^{-12}$ (w1) & TS (w1) & $F/10^{-12}$ (e1) 
& TS (e1) \\
(GeV) & (GeV) & (erg cm$^{-2}$ s$^{-1}$) & & (erg cm$^{-2}$ s$^{-1}$) 
& \\ 
\hline
0.13 & 0.1--0.2 & $<$6.2 & 0 & $<$5.5 & 0 \\
0.24 & 0.2--0.3 & $<$4.8 & 0 & $<$8.3 & 0 \\
0.42 & 0.3--0.6 & 3.5$\pm$1.7 & 10 & $<$1.9 & 0 \\
0.75 & 0.6--1.0 & 2.4$\pm$0.9 & 9 & $<$1.7 & 0 \\
1.33 & 1.0--1.8 & 1.2$\pm$0.5 & 5 & $<$0.5 & 0 \\
2.37 & 1.8--3.2 & $<$1.2 & 1 & $<$0.9 & 0 \\
4.22 & 3.2--5.6 & $<$1.2 & 2 & $<$0.9 & 1 \\
7.50 & 5.6--10.0 & $<$0.4 & 0 & $<$0.9 & 2 \\
13.30 & 10.0--17.8 & $<$0.2 & 0 & $<$0.2 & 0 \\
23.71 & 17.8--31.6 & $<$1.4 & 2 & $<$0.4 & 0 \\
42.17 & 31.6--56.2 & $<$1.8 & 2 & $<$0.7 & 0 \\
74.99 & 56.2--100.0 & $<$1.1 & 0 & $<$0.7 & 0 \\
\hline
\end{tabular}
\vskip 1mm
\footnotesize{Note: Flux upper limits are at a 95$\%$ confidence level.}
\end{table}

\subsection{Variability Analysis}
\label{subsec:va}

Since SS 433 and W50 region can be variable \citep{fab04}, we checked 
the possible variability of the excess emission at the w1 region. 
Following the procedure given in \citet{nol+12},
we calculated the variability 
index TS$_{var}$ in the 0.3--100 GeV energy range.
A total of 122 time bins were considered,  each containing 30-day data. 
In this case, TS$_{var}$ would be distributed as $\chi^{2}$ with 121 degrees 
of freedom if the source flux is constant.
We found TS$_{var}$= 103.0, which is lower than the threshold value of 160.1
for a variable source (at a 99\% confidence level; \citealt{nol+12}), 
indicating that no significant 
long-term variability was observed in the \gr\ emission.
\begin{figure}
\centering
\epsscale{1.0}
\plotone{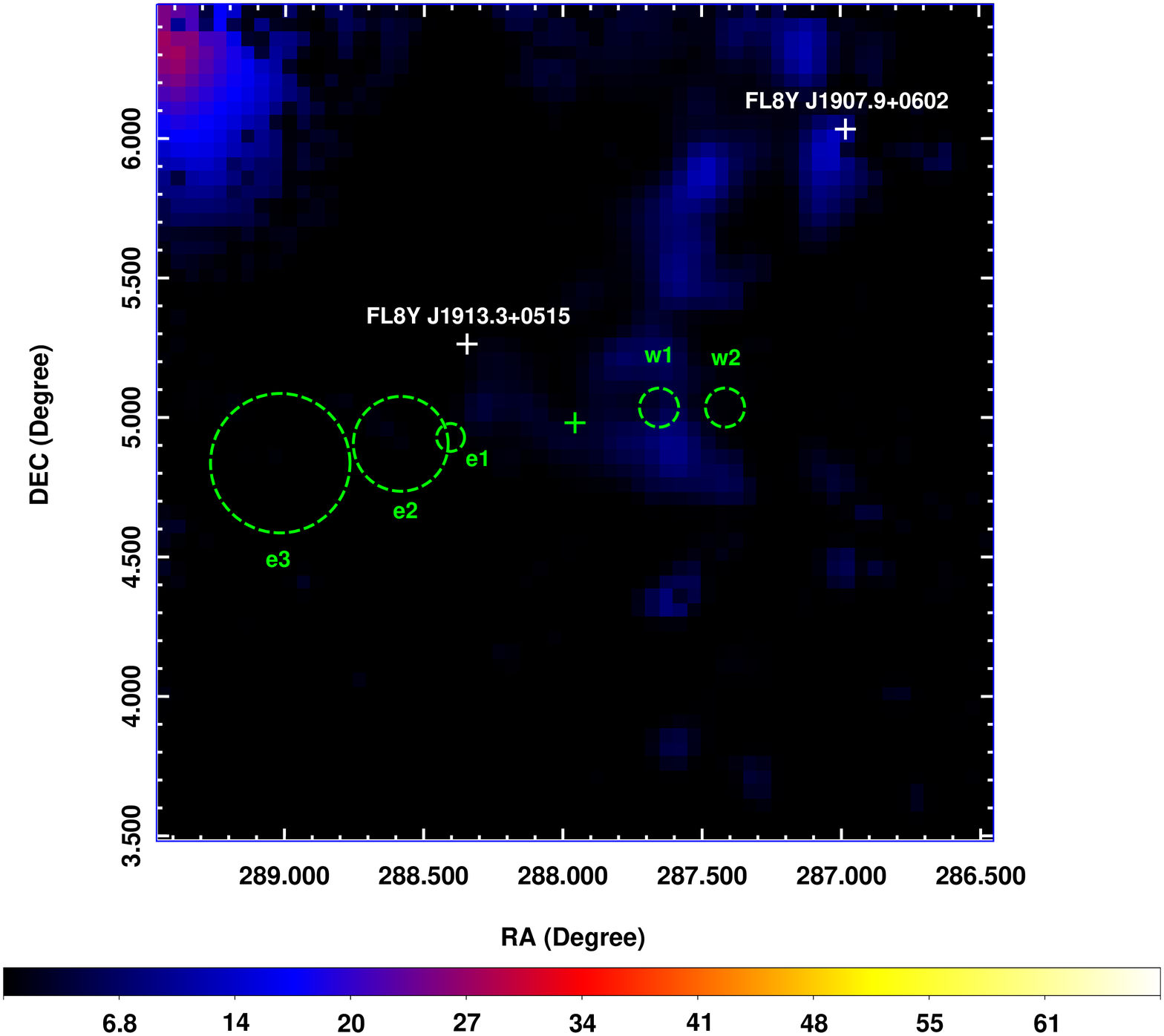}
\caption{$\mathrm{3^{o}\times3^{o}}$ TS residual map in 0.3--100 GeV band,
with the excess emission at w1 included as a point source in the source model
and removed.}
\label{fig:res}
\end{figure}

\subsection{Spatial Distribution Analysis}
\label{subsec:sda}

We examined whether the excess emission at the w1 region is a point source
or extended.
A point source with power-law emission and uniform disk models 
with power-law emission at the center of the w1 region were used.
The searched radius of the uniform disk ranges from 0.1 to 0.5 degree,
with a step of 0.1 degree. The TS$_{ext}$ value for the spatial extension 
was estimated from $-$2$\log(L_{ps}/L_{disk})$, where $L_{ps}$ and $L_{disk}$ 
are the maximum likelihood values for the point source and uniform disk model,
respectively. TS$_{ext}$ approximately is the square of the detection 
significance if the excess emission is extended \citep{lan+12}.
In this analysis, TS$_{ext}$ were smaller than 4, indicating
that no significant extendedness was present.

In addition, we also tested to include a point source
at w1 in the source model, and subtracted it in the resulting TS map. 
The excess emission was nearly totally removed in the residual TS map
(Figure~\ref{fig:res}).
This analysis again indicates that the excess emission can be considered
as a point source at w1.

\section{Discussion}
\label{sec:disc}

Analyzing ten years of the \fermi\ LAT Pass 8 data, we found significant 
\gr\ emission from the western interaction region in W50. 
The detection significance is $>$7$\sigma$ in the energy range of 0.3--100 GeV.
The location of the emission matches that of the X-ray bright spot and the
very recent HAWC detection
in the western lobe of W50 \citep{so97,abe+18}, making our results more
convincing than that reported in \citet{bor+15}. No variability or 
extendedness was found for the emission in our analysis. 
With a source distance of 5.5 kpc, 
the \gr\ luminosity is 1.8$\times$ 10$^{34}$ erg~s$^{-1}$. This luminosity 
is much lower than the kinetic energy power of the jets, which is at a level 
of $\sim$10$^{39}$ erg s$^{-1}$ \citep{mcs02}, indicating that there is 
sufficient energy to power the \gr\ emission.
\begin{figure*}
\centering
\epsscale{0.7}
\plotone{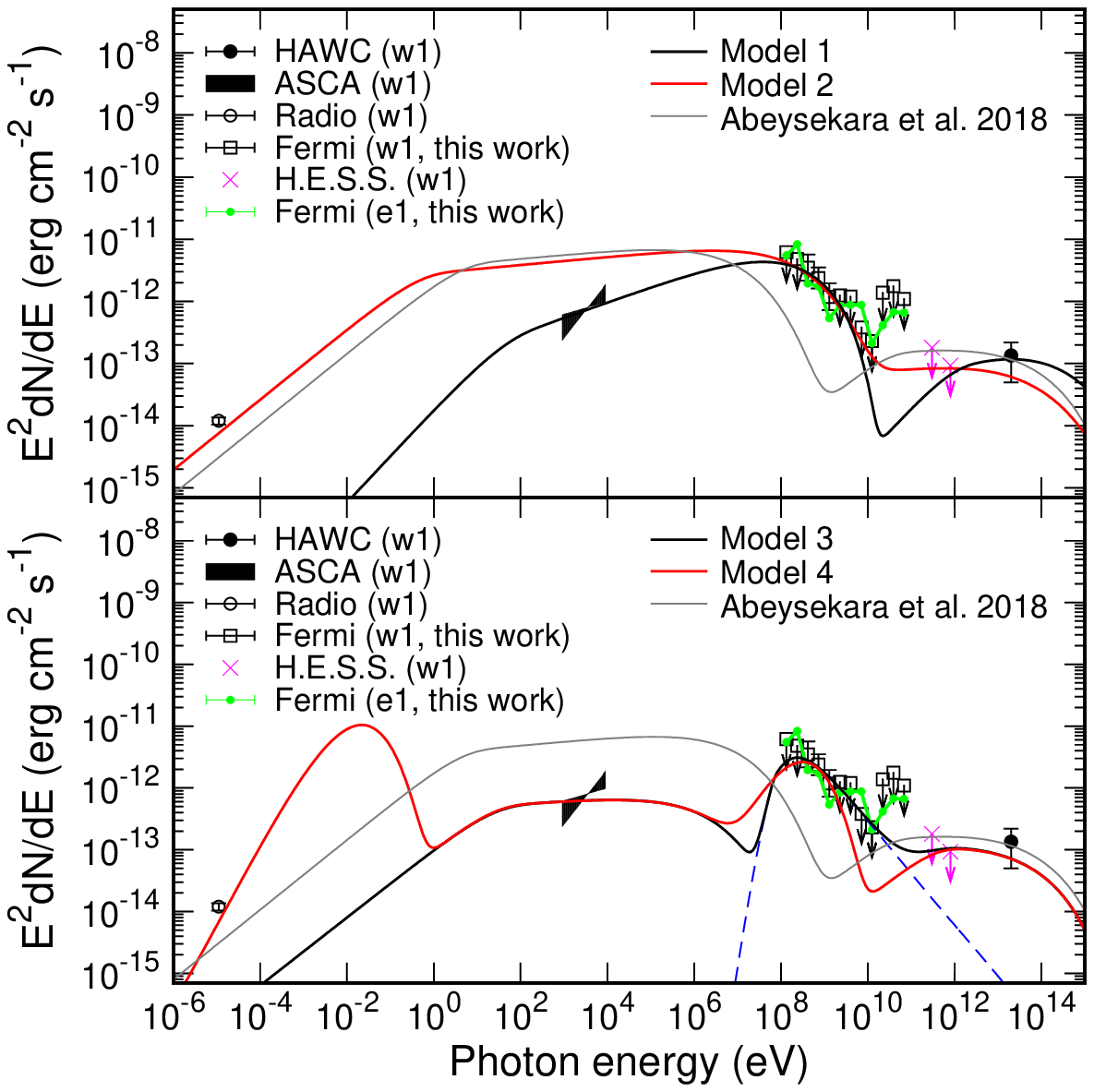}
\caption{\fermi\ LAT spectral energy distribution of the western emission 
region w1 in W50 (open squares). Also included are the HAWC flux at 20 TeV
(black dot; \citealt{abe+18}), H.E.S.S. 0.3 and 0.8 TeV upper limits 
(pink crosses; \citealt{magic18}), {\it ROSAT}
and {\it ASCA} flux (black bow-tie; \citealt{so97}), 
and radio fluxes estimated for w1 
(open circle; \citealt{gel+80}). 
For comparison, the leptonic model for the eastern emission 
region e1 given in \citet{abe+18} is shown as the grey curve 
and the GeV flux upper limits on the e1 region are shown as a green curve.
{\it Top panel:}
our leptonic Model 1 and 2 are indicated by the black and red curves 
respectively (see Section Discussion for details). {\it Bottom panel:} 
the hybrid Model 3 is indicated by the black curve,
in which the emission component from the hadronic process is shown
as the blue dashed curve, and Model 4 is indicated by the red curve, 
in which a relativistic Maxwellian component of electrons is included.}
\label{fig:sed}
\end{figure*}

The GeV \gr\ spectrum is very steep with $\Gamma\sim 6.0$ from 
300 MeV to 1.8 GeV, suggesting that
the detected emission should be an edge of an emission component. 
The flux is higher than that at $\sim$20 TeV from the HAWC observation and
that of their leptonic model at GeV energies (see Figure~\ref{fig:sed}; 
\citealt{abe+18}), raising a question about how to explain the \fermi\ 
measurements.
The broadband SED for the w1 region is thus constructed by including the HAWC
TeV \citep{abe+18}, X-ray \citep{so97}, and radio \citep{gel+80} measurements.  
In addition, we also include 
the VHE upper limits obtained with H.E.S.S. (deeper ones at $\sim$TeV 
energies;
\citealt{magic18}).

The leptonic model considered in \citet{abe+18} to explain the SED of the e1
region obviously cannot fit the detected GeV spectrum, lower by an order
of magnitude (see Figure~\ref{fig:sed}).
Here we also first construct a leptonic model, in which 
energetic electrons produce \grays\ by the IC process 
of scattering the CMB photons, and emissions from radio to X-ray or \gray\ 
via the synchrotron process.  
The radiation cooling, mainly due to synchrotron radiation, is included.
The current energy distribution of the electrons $dN/dE_e$ is described by 
$dN/dE_e\propto E_e^{-\alpha_e}{\rm exp}(-E_e/E_{c,e})$, where $E_{c,e}$ 
is the cutoff energy and $\alpha_e$ is the index of the power-law energy
distribution. The index becomes $\alpha_e+1$ if $E_e$ is greater than 
the break 
$E_{bre}=17\ (B/5\uG)^{-2}(t_{age}/30\ {\rm kyr})^{-1}$ TeV \citep{tan+08}.
In this case, the GeV and TeV \grays\ can not simultaneously originate 
from the IC process.
In order to match the observed data, the GeV \grays\ should be of synchrotron 
origin in a magnetic field of $B=5\uG$, while the TeV \grays\ be contributed 
by the IC radiation (black curve, Model 1 in the top panel of
Figure~\ref{fig:sed}). This requires the injected electron spectrum 
has a hard index 
$\alpha_e=1.5$ and a very large cutoff energy $E_{c,e}=28$ PeV.
The total energy in electrons with energy above 1 GeV is 
$W_{e}=0.7\times10^{46}$~erg, which is much lower than the total 
energy ($\sim 9\times 10^{51}$ erg) deposited in the jets during 
the timescale $t_{age}=30$ kyr.  However the model fails to explain 
the 2.6 GHz radio fluxes detected in the western region \citep{gel+80}. 

A model similar to that in \citet{abe+18} can be constructed, which
can fit the radio measurements (red curve, Model 2 in the top
panel Figure~\ref{fig:sed}). 
In this model, $\alpha_e=1.9$, $E_{c,e}=15$ PeV, and $B=20\uG$.
The problem of this model is that it gives an order of magnitude higher
X-ray fluxes than the observed. Because latter X-ray observations of the e1
region have shown higher fluxes \citep{sp99,bri+07} than that
reported from the {\it ROSAT} and {ASCA} observations \citep{so97}, we 
search for similar archival X-ray data, but unfortunately there is only one
{\it Chandra} imaging observation of the w2 region (Observation ID is 3843 and 
the exposure time is 71 ksec). The analysis of the data from this observation
gives a 0.6--6 keV luminosity of $\sim 2\times 10^{34}$\ erg\,s$^{-1}$,
only slightly higher than that given in \citet{so97}. Therefore there is
no observational evidence, even indirect one, that would support Model 2
at X-ray energies.

We test to include a hadronic process in which the \grays\ are generated 
in the collision between relativistic protons and the ambient gas.
The protons are also assumed to have a power-law energy 
distribution $dN/dE_p\propto E_p^{-\alpha_p}{\rm exp}(-E_p/3\ {\rm PeV})$.
We find that this population of protons also fail to simultaneously 
explain the GeV and TeV \gray\ fluxes.
Thus, we add a leptonic component with a fixed spectral index 
of $\alpha_e=2.0$ at injection and construct a lepton-hadron hybrid model.
In this case, the GeV \gray\ emission is due to the hadronic process 
with proton index $\alpha_{p}=2.9$, and the X-ray and TeV \grays\ are of
leptonic origin (black curve, Model 3 in the bottom panel
Figure~\ref{fig:sed}), requiring 
$E_{c,e}=2.5$ PeV, $B=5\uG$ and $W_{e}=4.8\times10^{46} $ erg.
This model also fails to explain the radio flux, and has a high total energy 
in protons $W_{p}=3\times10^{50}(n/1\ {\rm cm^{-3}})^{-1} $ erg,
which is comparable to the total jet energy even if the density of 
the ambient target gas is as high as $\sim1\ \rm cm^{-3}$ (see also 
arguments against a hadronic scenario in \citealt{abe+18}). 
Therefore this hybrid model may be excluded due to these problems.

Finally, we also test to replace the hadronic component in Model 3
with a relativistic thermal electron population (Model 4).
This model is based on the results in the particle-in-cell simulations of 
relativistic shocks \citep{s08}. In order to explain the resulting
particle distribution from the simulations, a model consisting of a 
relativistic 2D Maxwellian component plus a power-law component with an 
exponential cutoff is invoked.
This model has been applied to pulsar wind nebulae, explaining
their broadband spectra \citep[e.g.,][]{sla+12}.
In Model 4, we add a relativistic Maxwellian component 
$dN/dE_e\propto E_e^{2}{\rm exp}(-E_e/E_{peak})$, where $E_{peak}$ is 
the peak energy.  With this thermal component, we find that the broadband 
SED of the w1 region can be fitted (red curve in the bottom panel 
of Figure~\ref{fig:sed}), where reasonable parameters $E_{peak}=50$~GeV 
and $W_{e}^{max}=6.0\times10^{48}$~erg are obtained. However the model has
a steeply rising spectrum at radio frequencies. Such a spectrum is
not consistent with those obtained in radio observations of W50.
The observed spectral indices for W50 at radio frequencies are in a range
of $\sim 0.4$--0.7 (e.g., \citealt{dps86}), actually similar to 
the slope of the leptonic models (e.g., our Model 2).

In a summary, although GeV emission is clearly detected at the western region
w1 of SS 433/W50, matching well the X-ray and TeV detections,
it is hard to use a simple model to explain the current broadband
SED. One possibility is the X-ray flux of the w1 region was 
under-estimated. For the e1 region, a hard X-ray ($\sim$2--100 keV) power-law
component has been convincingly detected in different X-ray observations 
\citep{sp99, bri+07}. A clear detection of a similar component has not
been reported for the w1 
region, probably due to the lack of X-ray observations.
However, it should be noted that the jet emission is likely dominated
by a bremsstrahlung component (see, e.g., \citealt{mfm02}), and this thermal
component is also probably contained in the X-ray emission from the w1/e1 region
\citep{so97,bri+07}, which would reduce the X-ray flux considered in the SED.
In any case to clarify these, further X-ray observations of the w1 
region, particularly in the hard X-ray range, are needed.
In addition, the TeV fluxes at the w1 and e1 regions are approximately
equal \citep{abe+18}. 
Except that emission in 0.3--1.8 GeV from the w1 region
was weakly detected (with TS$\simeq$5--10), the flux upper limits
on the two sides in the other GeV bands are similar 
(see Table~\ref{tab:spectra} and Figure~\ref{fig:sed}).
If we consider that the physical properties of the two lobes of SS 433 should
have only small differences, the comparison may suggest that \fermi\ LAT might
be close to detecting the e1 region. This possibility can be checked when 
more LAT data are
collected. If this is the case, the model for the e1 region given in
\citet{abe+18} would also need to be modified.

As the current observational results have revealed fine features of a jet 
interaction region, detailed numerical simulations for investigating 
the interaction and radiation processes will help. It will identify 
different populations of high-energy particles. A model consisting of 
multiple particle components should definitely be able to explain the SED 
(for example, our Model 4, although the radio spectrum it produces is 
steeper than the observed). The possibility raised above
about why no GeV emission is seen in the eastern region of SS 433/W50
may also be investigated. We will be able to learn whether this deficit is
caused by very detailed differences in the interaction processes.

\acknowledgements
This research made use of the High Performance Computing Resource in the Core
Facility for Advanced Research Computing at Shanghai Astronomical Observatory.
This research was supported by the National Program on Key Research 
and Development Project (Grant No. 2016YFA0400804) and the National Natural 
Science Foundation of China (U1738131, 11633007). X.Z. and Y.C. acknowledge 
the supports from the National Program on Key Research and Development 
Projects 2018YFA0404204, 2017YFA0402600 and 2015CB857100 and from 
the NSFC under grants 11803011, 11773014 and 11851305.

\clearpage

\end{document}